\documentclass[aps,graphics]{revtex4}
\usepackage{epsfig}
\usepackage{graphicx}
\begin{document}
\title{
Total Teleportation of an Entangled State
}
\author{
Hai-Woong Lee\thanks{E-mail address: hwlee@laputa.kaist.ac.kr} }
\address{
Department of Physics, Korea Advanced Institute of Science and
Technology, Taejon 305-701,Korea }
\date{\today}

\begin{abstract}
We present a setup for a direct, total teleportation of a
single-particle entangled state. Our scheme consists of a
parametric down-conversion source, which emits a pair of entangled
photons, and a dual Bell measurement system with two beam
splitters and two pairs of detectors. An entangled state shared by
a pair of beams can be directly transferred to another pair of
beams with a 50\% probability of success. By a straightforward
generalization of the scheme, total teleportation of a
multi-particle entangled state can also be performed.

\vspace{0.25cm} PACS number(s): 03.67.-a, 03.65.Ta, 42.50.-p
\end{abstract}
\maketitle \vspace{0.5cm}

The recent progress and interest in quantum information science
stems largely from the realization that nonclassical features of
quantum systems can be utilized to achieve performance in
communications and computations that is superior to that based on
classical means\cite{book}. The nonlocal nature of quantum systems
arising from entanglement, in particular, has played a central
role in many recent exciting developments in this area of
research. Entanglement is an essential resource for many ingenious
applications in quantum information science such as quantum
teleportation\cite{Bennett,Bouwmeester} and quantum
cryptography\cite{Ekert,Jennewein} as well as for fundamental
studies in quantum mechanics related to Einstein-Podolsky-Rosen
(EPR) paradox\cite{Einstein} and Bell's theorem\cite{Bell}. In
Bell's test experiments and in quantum teleportation and
cryptography experiments, an important issue is whether one can
generate and manipulate entanglement in a controlled way.

In this paper we address ourselves to the issue of entanglement
transfer, more specifically, quantum teleportation of an unknown
entangled state. As pointed out by Bennett et al.\cite{Bennett} in
their original proposal for quantum teleportation, entanglement
can be transferred through teleportation of the state of one of
the entangled pair. This method, known as entanglement
swapping\cite{Zukowski,Bose,Pan}, provides partial teleportation
of entanglement in the sense that the teleported state is only a
part of the entangled state. We propose an alternative method in
which the entire entangled state is transferred directly from one
place to another. The method is demonstrated below for the case
when the state to be teleported is a single-particle entangled
state.

The teleportation scheme we propose is depicted in Fig. 1. The
state we wish to teleport is a single-photon entangled state of
the form $\alpha |1\rangle_{a_1} |0\rangle_{a_2} +\beta
|0\rangle_{a_1} |1\rangle_{a_2} $, where $\alpha$ and $\beta$ are
unknown except that $|\alpha|^2 +|\beta|^2 =1$. Here $|1\rangle$
and $|0\rangle$ refer to a one-photon state and a vacuum state,
respectively, and the subscripts $a_1$ and $a_2$ represent the
modes of photon in the beams $a_1$ and $a_2$, respectively. This
state may, for example, be generated when a photon is incident on
a beam splitter of unknown reflection and transmission
coefficients, $r=\alpha$, and $t=\beta$, as shown in the lower
left part of Fig. 1. At the source station we have a parametric
down-conversion source which emits a pair of entangled photons.
These photons emerge either through the ports $A_1$ and $B_1$ or
through $A_2$ and $B_2$. Such a source has been considered in past
studies of entanglement swapping\cite{Zukowski,Bose,Pan}. The
photons are thus in a two-photon entangled state
\begin{equation}
|\psi\rangle_{source} =\frac{1}{\sqrt{2}} (|1\rangle_{A_1}
|0\rangle _{A_2} |1\rangle_{B_1} |0\rangle_{B_2} +|0\rangle_{A_1}
|1\rangle _{A_2} |0\rangle_{B_1} |1\rangle_{B_2})
\end{equation}
The beams $A_1$ and $A_2$ are directed to Alice and the beams
$B_1$ and $B_2$ to Bob.

The state of the entire photon field before being detected by the
four detectors of Fig.1 is given by
\begin{equation}
|\psi\rangle_{in}=|\psi\rangle_{source} (\alpha |1\rangle_{a_1} |0
\rangle_{a_2} +\beta|0\rangle_{a_1} |1\rangle_{a_2})
\end{equation}
At Alice's station the beam $A_1$ from the source is combined via
a 50/50 beam splitter with the beam $a_1$ for a Bell measurement.
At the same time the beam $A_2$ is combined via another 50/50 beam
splitter with the beam $a_2$ for another Bell measurement.
Expressing the state $|\psi\rangle_{in}$ in the Bell basis, we can
write
\begin{eqnarray}
|\psi\rangle_{in}=\frac{1}{2\sqrt{2}}[(|\Phi^{(+)}_{A_1
a_1}\rangle |\Phi^{(+)}_{A_2 a_2}\rangle -|\Phi^{(-)}_{A_1
a_1}\rangle |\Phi^{(-)}_{A_2 a_2} \rangle)(\alpha|1\rangle_{B_1}
|0\rangle_{B_2} +\beta|0\rangle_{B_1}|1\rangle_{B_2} )\nonumber \\
+(|\Phi^{(-)}_{A_1 a_1}\rangle |\Phi^{(+)}_{A_2 a_2}\rangle
-|\Phi^{(+)}_ {A_1 a_1}\rangle |\Phi^{(-)}_{A_2
a_2}\rangle)(\alpha |1\rangle_{B_1}|0\rangle_{B_2}
-\beta|0\rangle_{B_1} |1\rangle_{B_2}) \nonumber \\
+(|\Psi^{(+)}_{A_1 a_1}\rangle |\Psi^{(+)}_{A_2 a_2}\rangle
-|\Psi^{(-)}_{A_1 a_1}\rangle|\Psi^{(-)}_{A_2 a_2}\rangle
)(\beta|1\rangle_{B_1} |0\rangle_{B_2} +\alpha|0\rangle_{B_1}
|1\rangle_{B_2}) \nonumber \\ +(|\Psi^{(-)}_{A_1
a_1}\rangle|\Psi^{(+)}_{A_2 a_2}\rangle -|\Psi^{(+)}_{A_1
a_1}\rangle |\Psi^{(-)}_{A_2 a_2}\rangle)(\beta|1\rangle _{B_1}
|0\rangle_{B_2} -\alpha|0\rangle_{B_1}|1\rangle_{B_2})]
\end{eqnarray}
where the Bell states are defined by
\begin{eqnarray}
|\Phi^{(\pm)}_{A_i a_i}\rangle =\frac{1}{\sqrt{2}}
(|1\rangle_{A_i} |1\rangle_{a_i} \pm |0\rangle_{A_i}
|0\rangle_{a_i}) \nonumber \\ |\Psi^{(\pm)}_{A_i a_i}\rangle=
\frac{1}{\sqrt{2}} (|1\rangle_{A_i}|0\rangle_{a_i} \pm
|0\rangle_{A_i} |1\rangle_{a_i})
\end{eqnarray}
If Alice's Bell measurement results in a simultaneous detection of
one photon each at the detector $D_{11}$ and $D_{21}$ and no
photon at $D_{12}$ and $D_{22}$ (corresponding to the input state
$|\Psi^{(+)}_{A_1 a_1}\rangle |\Psi^{(+)}_{A_2 a_2}\rangle$), or a
simultaneous detection of one photon each at $D_{12}$ and $D_{22}$
and no photon at $D_{11}$ and $D_{21}$ (corresponding to the input
state $|\Psi^{(-)}_{A_1 a_1}\rangle |\Psi^{(-)}_{A_2
a_2}\rangle$), then the state of the photon at Bob's station
reduces to $\beta|1\rangle_{B_1} |0\rangle_{B_2} +\alpha
|0\rangle_{B_1} |1\rangle_{B_2}$\cite{Comment}. If Bob is informed
of such a measurement result from Alice, he does not need to
perform any unitary transformation. He only needs to identify the
beam $B_2$ as a teleported version of the beam $a_1$ and the beam
$B_1$ as a teleported version of the beam $a_2$, and the
teleportation is achieved. If Alice's Bell measurement results in
a simultaneous detection of one photon each at $D_{11}$ and
$D_{22}$ and no photon at $D_{12}$ and $D_{21}$ (corresponding to
the input state $|\Psi^{(+)}_{A_1 a_1}\rangle |\Psi^{(-)}_{A_2
a_2}\rangle$), or a simultaneous detection of one photon each at
$D_{12}$ and $D_{21}$ and no photon at $D_{11}$ and $D_{22}$
(corresponding to the input state $|\Psi^{(-)}_{A_1 a_1}\rangle
|\Psi^{(+)}_{A_2 a_2}\rangle $), then the state of the photon at
Bob's station is $\beta|1\rangle_{B_1} |0\rangle_{B_2}-\alpha
|0\rangle_{B_1}|1\rangle_{B_2}$. An appropriate unitary
transformation can then be performed by Bob, and the teleportation
is achieved. Alice's Bell measurement, however, cannot distinguish
between the states $|\Phi^{(+)}_{A_1 a_1}\rangle |\Phi^{(+)}_{A_2
a_2}\rangle$, $|\Phi^{(-)}_{A_1 a_1}\rangle |\Phi^{(-)}_{A_2
a_2}\rangle$, $|\Phi^{(+)}_{A_1 a_1}\rangle |\Phi^{(-)}_{A_2
a_2}\rangle$, and $|\Phi^{(-)}_{A_1 a_1}\rangle |\Phi^{(+)}_{A_2
a_2}\rangle$, and therefore the probability of success for the
present scheme of entanglement teleportation is 50\%, the same as
that for the standard teleportation scheme\cite{Bennett}.

The teleportation experiment described above can be performed in
an "event-ready" fashion by using the technique of entanglement
swapping, as suggested by Zukowski et al.\cite{Zukowski} For this
purpose we employ at the source station the setup shown in Fig.2.
It consists of two independent sources, two 50/50 beam splitters
and two pairs of detectors. The two sources each emit one pair of
entangled photons in the state $\frac{1}{\sqrt{2}}
(|1\rangle_{A_1} |0\rangle_{B_2} |1\rangle_{G_1} |0\rangle_{H_2}
+|0\rangle_{A_1} |1\rangle_{B_2} |0\rangle_{G_1} |1\rangle_{H_2})
\frac{1}{\sqrt{2}}(|1\rangle_{A_2} |0\rangle_{B_1} |1\rangle_{G_2}
|0\rangle_{H_1} +|0\rangle_{A_2} |1\rangle_{B_1} |0\rangle_{G_2}
|1\rangle_{H_1})$. Detection of one photon each at $D_{G_1}$ and
$D_{H_1}$ and no photon at $D_{G_2}$ and $D_{H_2}$, or detection
of one photon each at $D_{G_2}$ and $D_{H_2}$ and no photon at
$D_{G_1}$ and $D_{H_1}$, would indicate that the state collapses
into $|\Psi^{(\pm)}_{G_1 G_2}\rangle |\Psi^{(\pm)}_{H_1
H_2}\rangle \frac{1}{\sqrt{2}} (|1\rangle_{A_1} |0\rangle_{A_2}
|1\rangle_{B_1} |0\rangle_{B_2}+ |0\rangle_{A_1} |1\rangle_{A_2}
|0\rangle_{B_1} |1\rangle_{B_2})$. Thus, the state of the photons
emerging eventually from the source subjected to either of the
above detection results becomes exactly the state
$|\psi\rangle_{source}$ of Eq.(1). The detectors $D_{11}$,
$D_{12}$, $D_{21}$, and $D_{22}$ can be activated when the
detectors $D_{G_1}$, $D_{G_2}$, $D_{H_1}$, and $D_{H_2}$ register
either of the above results.

The proposed scheme of teleporting single-particle entanglement
can be generalized to the case of multi-particle entanglement in a
straightforward way. Let us consider teleportation of an
N-particle entangled state of the form
\begin{equation}
\alpha|1\rangle_{a_1}|0\rangle_{a_2} |1\rangle_{b_1}
|0\rangle_{b_2} \cdots |1\rangle_{p_1}|0\rangle_{p_2}+\beta
|0\rangle_{a_1}|1\rangle_{a_2} |0\rangle_{b_1}|1\rangle_{b_2}
\cdots |0\rangle_{p_1} |1\rangle_{p_2}
\end{equation}
The quantum channel that links Alice and Bob is provided by an
(N+1)-particle entangled state
\begin{equation}
\frac{1}{\sqrt{2}}(|1\rangle_{A_1} |0\rangle_{A_2} |1\rangle_{B_1}
|0\rangle_{B_2} \cdots |1\rangle_{P_1} |0\rangle_{P_2}
|1\rangle_{Q_2} |0\rangle_{Q_2}+|0\rangle_{A_1} |1\rangle_{A_2}
|0\rangle_{B_1} |1\rangle_{B_2} \cdots |0\rangle_{P_1}
|1\rangle_{P_2} |0\rangle_{Q_1} |1\rangle_{Q_2})
\end{equation}
In Eqs.(5) and (6) the letters $p$ and $P$ serve to identify the
N-th particle and the letter $Q$ the (N+1)-th particle. The
(N+1)-particle entangled state of Eq. (6) is generated at the
source station. The beams $A_1$ and $A_2$, along with the beams
containing the unknown entangled state, are sent to Alice, while
all other beams are sent to Bob. With (N+1) beam splitters and
(N+1) pairs of detectors, Alice then makes a Bell measurement on
the pairs ($A_1, a_1$), ($A_2 ,a_2$), ($b_1 ,b_2$),($c_1 ,c_2$),
$\cdots$, ($p_1 ,p_2$), separately.

The photon field state given by the product of Eqs. (5) and (6)
can be expressed in the Bell basis. Straightforward algebra shows
that the photon field state expressed in the Bell basis can be
divided into four groups, each associated with a different field
state at Bob's station. Two of the four groups contain products of
the states of the form $|\Phi^{(\pm)}_{A_1 a_1}\rangle
|\Phi^{(\pm)}_{A_2 a_2}\rangle$, and are not of much use, because
these states cannot be distinguished from each other by Alice's
Bell measurement. The remaining two groups can be expressed as
\begin{eqnarray}
\sum_{even} (\pm) |\Psi^{(\pm)}_{A_1 a_1}\rangle
|\Psi^{(\pm)}_{A_2 a_2}\rangle |\Psi^{(\pm)}_{b_1 b_2}\rangle
|\Psi^{(\pm)}_{c_1 c_2}\rangle \cdots |\Psi^{(\pm)}_{p_1 p_2}
\rangle(\beta|1\rangle_{B_1} |0\rangle_{B_2} |1\rangle_{C_1}
|0\rangle_{C_2} \cdots |1\rangle_{Q_1}|0\rangle_{Q_2} \nonumber \\
+\alpha |0\rangle_{B_1} |1\rangle_{B_2} |0\rangle_{C_1}
|1\rangle_{C_2} \cdots |0\rangle_{Q_1} |1\rangle_{Q_2}) \nonumber
\\ +\sum_{odd} (\pm) |\Psi^{(\pm)}_{A_1 a_1}\rangle
|\Psi^{(\pm)}_{A_2 a_2} \rangle |\Psi^{(\pm)}_{b_1 b_2}\rangle
|\Psi^{(\pm)}_{c_1 c_2} \rangle \cdots |\Psi^{(\pm)}_{p_1 p_2}
\rangle (\beta |1\rangle_{B_1} |0\rangle_{B_2} |1\rangle_{C_1}
|0\rangle_{C_2} \cdots |1\rangle_{Q_1} |0\rangle_{Q_2} \nonumber
\\ -\alpha |0\rangle_{B_1} |1\rangle_{B_2} |0\rangle_{C_1}
|1\rangle_{C_2} \cdots |0\rangle_{Q_1} |1\rangle_{Q_2})
\end{eqnarray}
where $\sum_{even}$ [$\sum_{odd}$] denotes that, out of all
possible combinations of $|\Psi^{(+)}\rangle$'s and $|\Psi^{(-)}
\rangle$'s in the product of (N+1) Bell states, only those for
which the total number of $|\Psi^{(-)}\rangle$'s is even [odd]
should be included in the summation, and $(\pm)$ immediately
following the summation sign indicates that some of the terms in
the summation are added and some others subtracted.

It is important to note that different combinations of the product
of (N+1) $|\Psi^{(\pm)}\rangle$'s have different combinations of
detectors registering a photon. Thus, when the (N+1) pairs of
detectors register a particular combination of detection
coincidence, we know which combination of the product of
$|\Psi^{(\pm)}\rangle$'s it corresponds to and consequently which
state Bob has in the beams ($B_1 ,B_2 ,C_1 ,C_2 ,\cdots,Q_1 ,Q_2
$). An appropriate unitary transformation performed by Bob then
completes the teleportation process. Since the state given by
Eq.(7) represents one half of the entire initial photon state, the
probability of success for teleportation of an N-particle
entangled state is 50\%.

Let us consider the case N=2. This case corresponds to teleporting
a two-particle entangled state using three-particle entanglement
and has recently been considered by Shi et al.\cite{Shi} and by
Marinatto and Weber\cite{Marinatto}. Our scheme for teleporting a
two-particle entangled state $\alpha |1\rangle_{a_1}
|0\rangle_{a_2} |1\rangle_{b_1} |0\rangle_{b_2} +\beta
|0\rangle_{a_1} |1\rangle_{a_2} |0\rangle_{b_1} |1\rangle_{b_2}$
is shown in Fig.3. Here the quantum channel should be provided by
a three-particle Greenberger-Horne-Zeilinger (GHZ)
state\cite{Greenberger} $\frac{1}{\sqrt{2}}(|1\rangle _{A_1}
|0\rangle_{A_2} |1\rangle_{B_1} |0\rangle_{B_2} |1\rangle_{C_1}
|0\rangle_{C_2} +|0\rangle_{A_1} |1\rangle_{A_2} |0\rangle_{B_1}
|1\rangle_{B_2} |0\rangle_{C_1} |1\rangle_{C_2})$. This type of
GHZ state can be generated using the method of Zeilinger et
al.\cite{Zeilinger} In this situation, a coincidence detection of
a single photon by the detectors ($D_1$, $D_3$, and $D_5$), or
($D_1$, $D_4$, and $D_6$), or ($D_2$, $D_3$, and $D_6$), or
($D_2$, $D_4$, and $D_5$), collapses the state at Bob's station
into $\beta|1\rangle_{B_1} |0\rangle_{B_2} |1\rangle_{C_1}
|0\rangle_{C_2}+ \alpha|0\rangle_{B_1} |1\rangle_{B_2}
|0\rangle_{C_1} |1\rangle_{C_2}$. This occurs with a 25\%
probability. On the other hand, a coincidence detection of a
single photon by the detectors ($D_1$, $D_3$, and $D_6$), or
($D_1$, $D_4$, and $D_5$), or ($D_2$, $D_3$, and $D_5$), or
($D_2$, $D_4$, and $D_6$), collapses the state at Bob's station
into $\beta|1\rangle_{B_1} |0\rangle_{B_2} |1\rangle_{C_1}
|0\rangle_{C_2} -\alpha |0\rangle_{B_1} |1\rangle_{B_2}
|0\rangle_{C_1} |1\rangle_{C_2}$. This occurs also with a 25\%
probability. The case N=3 corresponds to teleportation of an
unknown GHZ-type entangled state $\alpha |1\rangle_{a_1}
|0\rangle_{a_2} |1\rangle_{b_1} |0\rangle_{b_2} |1\rangle_{c_1}
|0\rangle_{c_2} +\beta |0\rangle_{a_1} |1\rangle_{a_2}
|0\rangle_{b_1} |1\rangle_{b_2} |0\rangle_{c_1} |1\rangle_{c_2}$.
The quantum channel in this case is provided by a four-particle
entangled state $\frac{1}{\sqrt{2}} (|1\rangle_{A_1}
|0\rangle_{A_2} |1\rangle_{B_1} |0\rangle_{B_2} |1\rangle_{C_1}
|0\rangle_{C_2} |1\rangle_{E_1} |0\rangle_{E_2}+ |0\rangle_{A_1}
|1\rangle_{A_2} |0\rangle_{B_1} |1\rangle_{B_2} |0\rangle_{C_1}
|1\rangle_{C_2} |0\rangle_{E_1} |1\rangle_{E_2})$, which can be
generated, for example, using the method of Bose et al.\cite{Bose}
which is a generalized version of the method of Zukowski et
al.\cite{Zukowski} or of Zeilinger et al.\cite{Zeilinger}.

Bennett et al.\cite{Bennett}, in their original proposal for
quantum teleportation, have already noted that teleportation can
also be achieved for entangled states. Let particles $A$ and $B$
be an entangled pair produced at the source station and sent to
Alice and Bob, respectively, and let particles $a$ and $b$ be a
pair in an unknown entangled state. If Alice makes a Bell
measurement on the particles $A$ and $a$, the state of particle
$a$ can be teleported to particle $B$ at Bob's station. In
addition, Bennett et al. showed that the particles $B$ and $b$ are
entangled exactly the same way the particles $a$ and $b$ were
before teleportation. This is entanglement
swapping\cite{Zukowski,Bose,Pan} and provides a way of teleporting
an entangled state via teleportation of the state of only one
particle of the entangled pair. Teleportation of an unknown
single-particle entangled state using this method has been
discussed in our earlier publication\cite{Lee}. We note, however,
that, if Bob is to have the desired entangled pair in his
possession, not only the particle $B$ should be sent to him from
the source station, but also the particle $b$ should be arranged
to be sent to him. In contrast, the scheme described here offers a
way for a direct, total teleportation of an entangled state. The
entangled state shared by the two beams $a_1$ and $a_2$ of Fig. 1
is directly teleported to the beams $B_1$ and $B_2$, which were
sent to Bob from the source station. The beams $a_1$ and $a_2$ in
an unknown entangled state are both subjected to Alice's Bell
measurement and neither needs to be sent to Bob. We also note that
the present scheme allows teleportation of an entangled state with
only three photons, the same number necessary to teleport a qubit
of states in the standard teleportation scheme.

One may argue that the necessity of sending the particle $b$ to
Bob in the standard teleportation scheme can be avoided if the
particle $b$ is subjected to a second Bell measurement at Alice's
station. It then is necessary that the source station generates
two entangled pairs, say ($A_1 ,B_1$) and ($A_2 ,B_2 $). The
particles $A_1$ and $A_2$ are sent to Alice and the beams $B_1$
and $B_2$ to Bob. Alice makes two successive Bell measurements,
first on the pair ($A_1$ and $a$) and second on the pair ($A_2$
and $b$). the entanglement can be transferred this way from the
particles $a$ and $b$ to the particles $B_1$ and $B_2$. This
method, however, has only a 25\% probability of success, because
each Bell measurement is associated with a 50\% probability of
success. Furthermore, the method may not succeed if the quantum
channel is noisy, because the initial state for the second Bell
measurement is no longer pure\cite{Lee2}. Apparently, the present
scheme does not suffer from this difficulty.

Finally, technical difficulties with the present scheme should be
pointed out. Since the present scheme utilizes single-particle
entanglement and its multi-particle generalizations, it requires
production, maintenance, and detection of photons at a single
photon level. In particular, the Bell measurements required in the
present scheme consist of identifying particular combinations of
coincidence detection of a single photon among (N+1) detectors.
Although simple in principle, such Bell measurements require the
highest degree of sensitivity on detectors, because detectors
should be capable of distinguishing between no photon, one photon
and two photons.

This research was supported by the Brain Korea 21 Project of the
Korean Ministry of Education and by Korea Research Foundation. The
author wishes to thank Professor J. Kim for stimulating
discussions and Dr. D. H. Kwon for technical assistance in
figures.

\begin{figure}
\includegraphics[scale=.5]{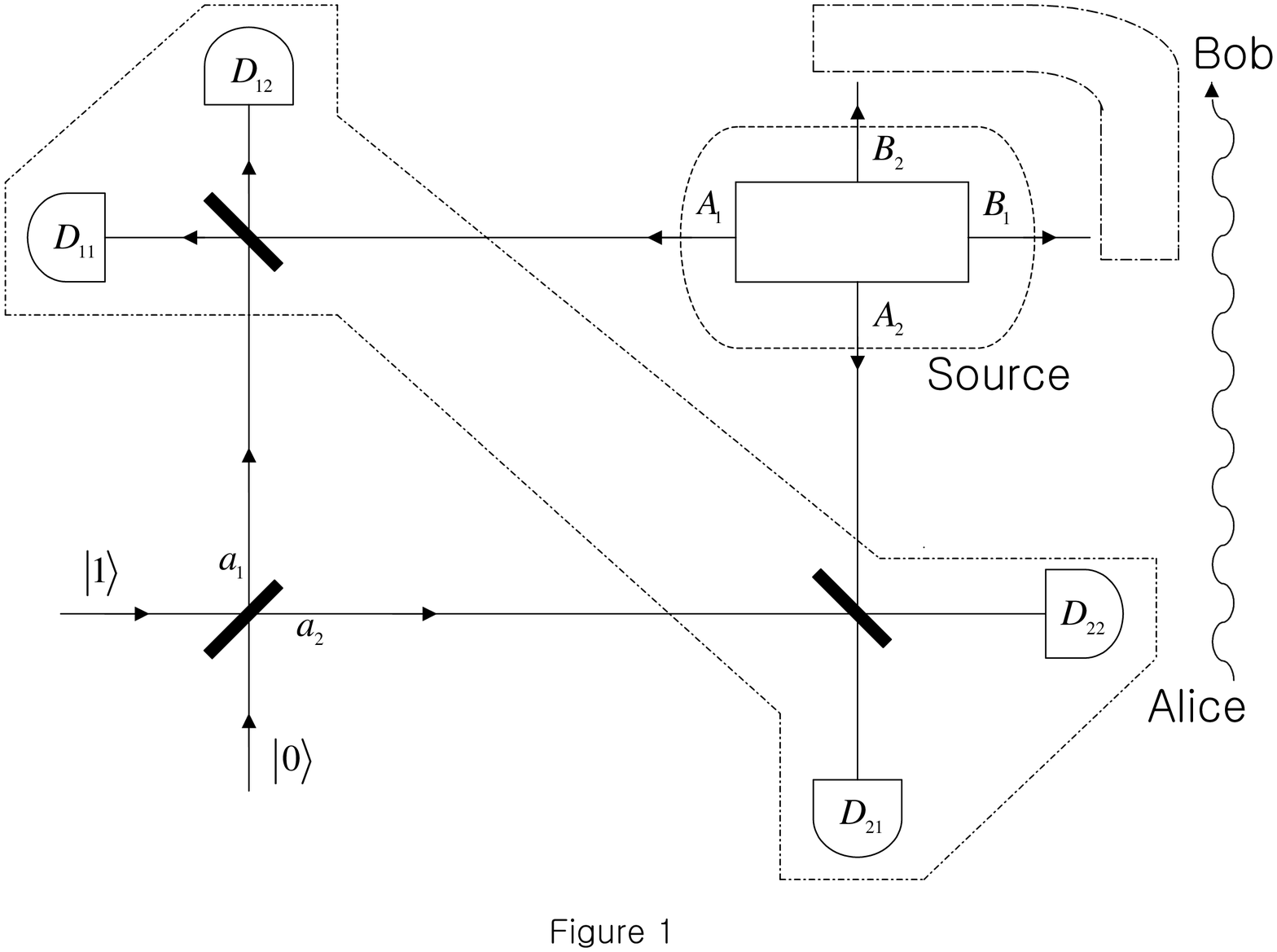}
\caption{Teleportation of a single-particle entangled state. An
unknown entangled state in the beams $a_1$ and $a_2$ is
teleported to the state in the beams $B_1$ and $B_2$. The wavy
line represents a classical communication channel.}
\end{figure}
\begin{figure}
\includegraphics[scale=.5]{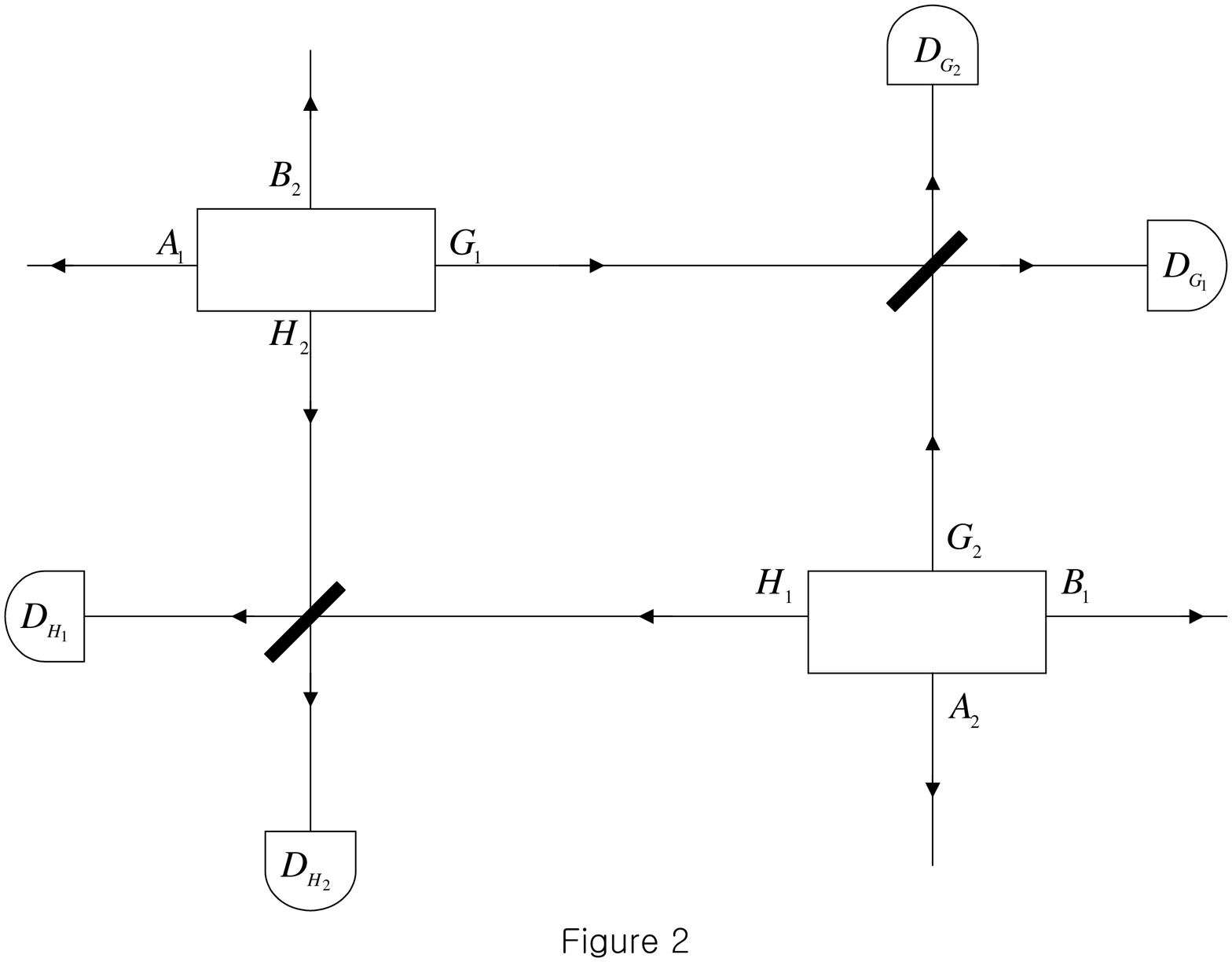}
\caption{Setup at the source station for an "event-ready"
experiment.}
\end{figure}
\begin{figure}
\includegraphics[scale=.5]{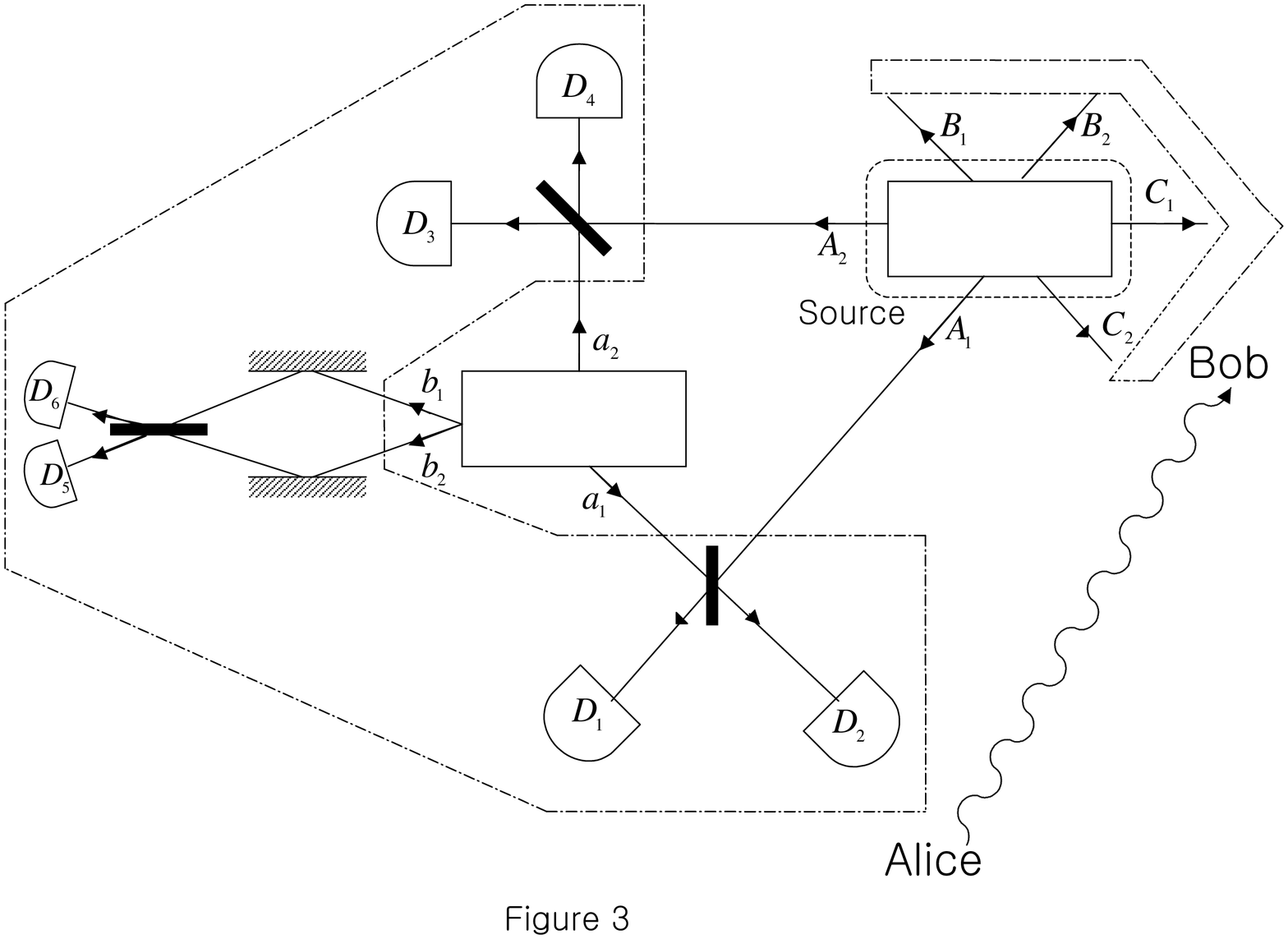}
\caption{Teleportation of a two-particle entangled state. An
unknown entangled state in the beams $a_1$, $a_2$, $b_1$, and
$b_2$ is teleported to the state in the beams $B_1$, $B_2$,
$C_1$, and $C_2$. The wavy line represents a classical
communication channel.}
\end{figure}


\begin{references}
\bibitem{Comment} Here and throughout this paper, we assume that
beam splitters are equipped with appropriate phase shifters such
that two output modes are a symmetric and antisymmetric
combination of two input modes.
\end{references}
\end{document}